\documentclass[preprint,preprintnumbers,amsmath,amssymb]{revtex4}

\usepackage{graphicx}
\usepackage{dcolumn}
\usepackage{bm}

\begin{document}

\title{Quantitative analysis of privatization}
\author{M. Vahabi, G. R. Jafari \\
Department of Physics, Shahid Beheshti University, Evin, Tehran
19839, Iran}

\begin{abstract}

In recent years, the economic policy of privatization, which is
defined as the transfer of property or responsibility from public
sector to private sector, is one of the global phenomenon that
increases use of markets to allocate resources. One important
motivation for privatization is to help develop factor and product
markets, as well as security markets. Progress in privatization is
correlated with improvements in perceived political and investment
risk. Many emerging countries have gradually reduced their political
risks during the course of sustained privatization. In fact, most
risk resolution seems to take place as privatization proceeds to its
later stage. Alternative benefits of privatization are improved risk
sharing and increased liquidity and activity of the market. One of
the main methods to develop privatization is entering a new stock to
the markets for arising competition. However, attention to the
capability of the markets to accept a new stock is substantial.
Without considering the above statement, it is possible to reduce
the market's efficiency. In other words, introduction of a new stock
to the market usually decreases the stage of development and
activity and increases the risk. Based on complexity theory, we
quantify how the following factors: stage of development, activity,
risk and investment horizons play roles in the privatization.

\end{abstract}
\maketitle

\section{Introduction}

In recent years, economics and finance have been at the focus of
many researchers in various fields. Among these researchers,
physicists, Mathematicians and engineers attempt to apply existing
knowledge from mathematics' approaches to economic problems
\cite{McCauley,Stanley,Sornette,Roehner,Chakrabarti,Bouchaud,Ball,Samanidou,Friedrich}.
The aim is to characterize the statistical properties of given
markets time series with the hope to provide useful information to
create new models able to reproduce experimental facts.

In economics, a financial market is a mechanism that allows people
to participate as investors and easily buy and sell financial
securities, commodities, etc at low transaction costs and at prices.
In fact, producers have obtained a feed back from consumers. This is
because consumers are their investors and their benefits lies in the
development and success of the markets. In finance, financial
markets facilitate the raising of capital (in the capital markets),
the transfer of risk (in the derivatives markets) and international
trade (in the currency markets). Among various economic problems,
privatization has been an interesting one for many researchers
\cite{Bortolotti1,Bortolotti2,Bortolotti3,Bortolotti4,Brada,Vining,Shleifer,Vishny1,Kay,Vickers,Bishop,Vishny2}.
Nowadays, the economic policy of privatization, which is defined as
the transfer of property or responsibility from government to
business, is a matter of interest in many countries. We assume that
the goal of government is to promote efficiency. Indeed, it is now
quite difficult to find a country that has not embarked on a program
to involve the private sector in their management, ownership, and
financing. Even if privatization processes seem to pursue a common
global trend, the extent of divestiture varies greatly across
countries. In some countries, governments have followed a consistent
and continuous privatization policy as a part of wider reform
packages, while in some others, it has been sporadic and
small-scaled \cite{Bortolotti3}. However, there is no doubt that
privatization has had a major impact on capital markets and trading
volumes. Privatization can range from a simple contract with a
private vendor to the sale of a public asset. There are many reasons
why governments turn to privatization. Cost reduction is one
motivation for privatization. The desire to transfer risk from the
public sector to the private sector can lead to privatization. A
higher level of service and an absence of expertise within the
governmental unit can also be another reasons. The time frame with
which a project needs to be completed could also factor in the
decision for privatization. A final potential reason for
privatization is the flexibility provided by the private sector. One
of the methods for privatization is share issue privatization. In
this method, the government sells shares of the government run
company which can then be traded on various stock markets, though a
developed secondary market is necessary \cite{commission}. Indeed,
financial market development is mentioned as one of the primary
objectives of privatization. A remarkable wealth of evidence shows
the correlation between financial market development and
privatization. Yet, stock markets develop also in the absence of
privatization \cite{Bortolotti2}. Proponents of privatization
believe that private market actors can more efficiently deliver any
good or service that government can provide. Privatization
proponents' faith in the market is philosophically based on an
economic principle of competition: where there is a profit to be
made, competition will inevitably arise, and that competition will
inevitably draw prices down while increasing efficiency and quality.
However, some would point out that privatizing certain functions of
government might hamper coordination, and charge firms with
specialized and limited capabilities to perform functions which they
are not suited for.

It has been clear in the transition economies that the success of
the privatization program depends on the strength of the markets
within the same country, and vice versa. Thus, the impact of
privatization will differ across countries depending on the strength
of the existing private sector. Similarly, the evidence suggests
that the effectiveness of privatization depends on institutional
factors, such as the protection of investors. However, privatization
can also stimulate the development of institutions that improve the
operations of markets. A key decision to be made by the privatizing
government is the method through which the state-owned asset is
transferred to private ownership. This decision is difficult
because, in addition to the economic factors such as valuing the
assets, privatization is generally part of an ongoing, highly
politicized process. Some of the factors that influence the
privatization method include: (1) the history of the asset's
ownership, (2) the need to pay off important interest groups in the
privatization, (3) the capital market conditions and existing
institutional framework for corporate governance in the country, (4)
the sophistication of potential investors, and, (5) the government's
willingness to let foreigners own divested assets \cite{william}.
The complexity of the goals of the process means that different
countries have used many different methods for privatizing various
types of assets. Although financial economists have learned much
about selling assets in well-developed capital markets, we still
have a limited understanding of the determinants and the
implications of the privatization method for state-owned assets.
Theoreticians have modeled some aspects of the privatization process
but, to be tractable, their models must ignore important factors.
Empirical evidence on the determinants of privatization is also
limited by the complexity of the goals of the privatization process.

Progress in privatization is correlated with improvements in
perceived political risk. These gains tend to be gradual over the
privatization period and are significantly larger in privatizing
countries than in non-privatizing countries, suggesting that the
resolution of such risk is endogenous to the privatization process.
Changes in political risk in general tend to have a strong effect on
local stock market development and excess returns in emerging
economies, suggesting that political risk is a priced factor. The
resolution of political risk resulting from successful privatization
has been an important source for the rapid growth of stock markets
in emerging economies. The recent wave of privatization sales in
developing countries should have altered the perceived political
risks of these countries considerably, especially if governments
have successfully implemented the announced privatization plans.
Such shifts in political risk tend to affect the attractiveness of
equity investments and are therefore related to stock market
development. Many emerging countries have gradually reduced their
political risks during the course of sustained privatization. In
fact, most risk resolution seems to take place as privatization
proceeds to its later stage. The known benefits of privatization are
reduction in public debt, improved incentives and efficiency, and
better access to capital \cite{perotti}.

We relate how privatization depends on stock market development.
Furthermore, liquidity, rather than capitalization, provides
incentives for information acquisition to financial analysts.  Based
on recent researches, concepts of high activity and degree of
development and low risk of the markets have been defined
\cite{Payam,Matteo1,Matteo2,jafari1,mahsa}. Some reports indicate
that \cite{jafari1,mahsa} Level Crossing (LC) and Hurst exponent
show remarkable differences between developed and emerging markets.
Level Crossing analysis is very sensitive to correlation when the
time series is shuffled and to probability density functions (PDF)
with fat tails when the time series is surrogated.

\section{Level Crossing Analysis}

Let us consider a time series $\{p(t)\}$, of price index with length
$n$, and the price returns $r(t)$ which is defined by $r(t) = \ln
p(t+1)-\ln p(t)$.

Let for a typical time interval $T$, $\nu_{\alpha}^{+}$ denotes the
number of positive difference crossings $r(t)- \bar r = \alpha$ in
time $t$ (see figure $1$ ) and let the mean value for all the time
intervals be $N_{\alpha}^{+}(T)$ where \cite{Tabar1}:

\begin{equation}
N_{\alpha}^{+}(T)=E[n_{\alpha}^{+}(T)].
\end{equation}

In other words, $\nu_{\alpha}^{+}$ is the average frequency of
positive slope crossings of the level $\alpha$.

For the homogeneous process, if we take a second interval of $T$
immediately following the first we shall obtain the same result, and
for the two intervals together we shall therefore obtain
\cite{Tabar1}
\begin{equation}
N_{\alpha}^{+}(2T)=2N_{\alpha}^{+}(T),
\end{equation}
from which it follows that, for a homogeneous process, the average
number of crossings is proportional to the time interval $T$. Hence
\begin{equation}
N_{\alpha}^{+}(T)\propto T,
\end{equation}
or
\begin{equation}
N_{\alpha}^{+}(T)=\nu^{+}_{\alpha} T.
\end{equation}

which $\nu_{\alpha}^{+}$ is the average frequency of positive slope
crossings of the level $r(t)- \bar r = \alpha$. We now consider how
the frequency parameter $\nu_{\alpha}^{+}$ can be deduced from the
underlying probability distributions for $r(t)- \bar r$. Consider a
time scale $\Delta t$ of a typical sample function, if $r(t)-\bar r
< \alpha$ at time $t$ and $r(t)- \bar r > \alpha$ at $t+\Delta t$ or
alternatively the changes in r(t) is positive in the time interval
$\Delta t$, there will be a positive crossing of $r(t)-\bar r
=\alpha$,

\begin{equation}
r(t)- \bar r < \alpha \hspace{.6cm} and \hspace {.6cm}
\frac{\Delta(r(t)-\bar r)}{\Delta t}> \frac{\alpha-(r(t) - \bar r)
}{\Delta t}.
\end{equation}


\begin{figure}[t]
\includegraphics[width=12cm,height=10cm,angle=0]{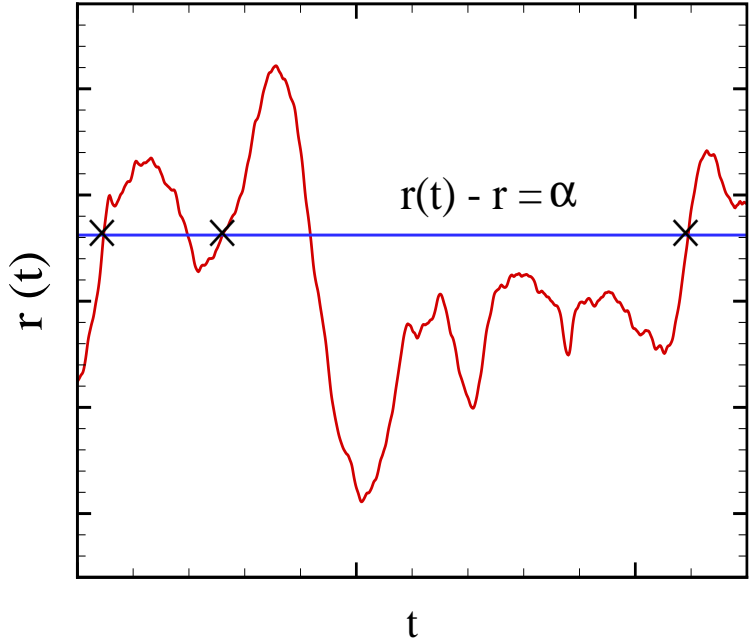}
\caption{Schematic positive slope crossings in a fixed level,
$\alpha$.}\label{fig1}
\end{figure}

\begin{figure}[t]
\includegraphics[width=17cm,height=16cm,angle=0]{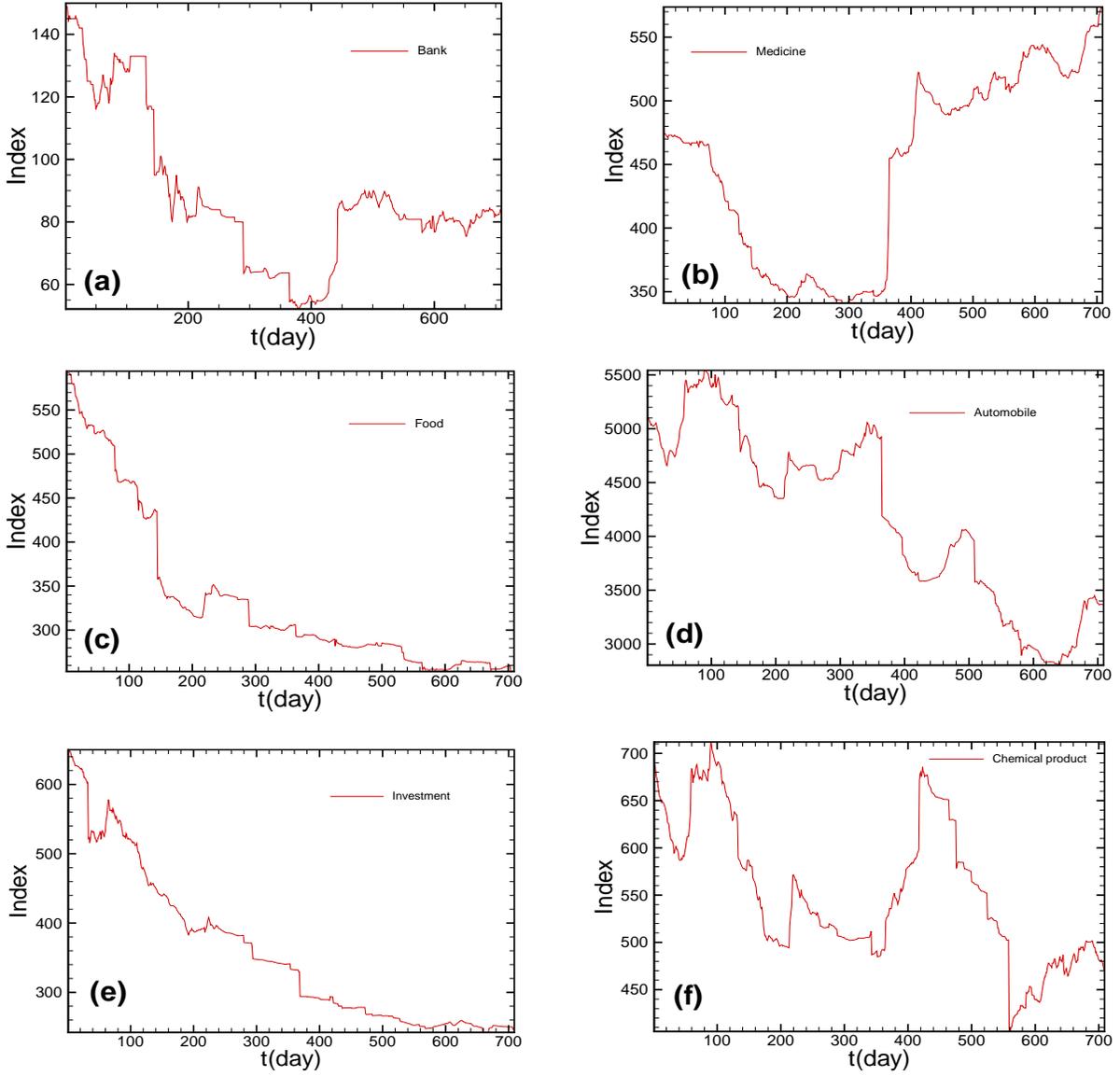}
\caption{Indices history (Jan. 2005 - Mar. 2008) of some TEPIX
 subgroups (a) Bank, (b) Medicine, (c) Food, (d) Automobile, (e) Investment and (f) Chemical products.}\label{fig2}
\end{figure}

\begin{figure}[t]
\includegraphics[width=12cm,height=10cm,angle=0]{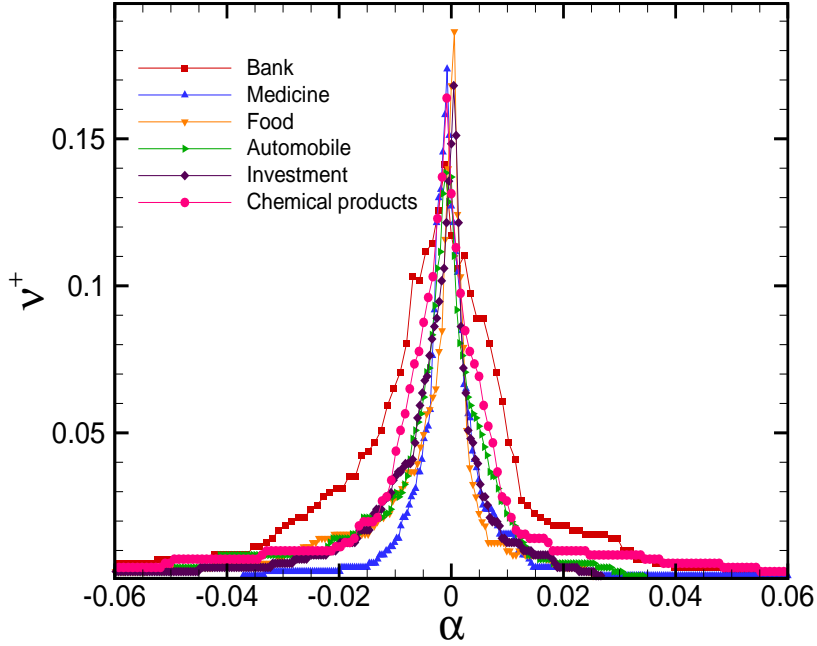}
\caption{Comparison of the positive Level Crossings of some TEPIX
subgroup indices.}\label{fig3}
\end{figure}

\begin{figure}[t]
\includegraphics[width=12cm,height=10cm,angle=0]{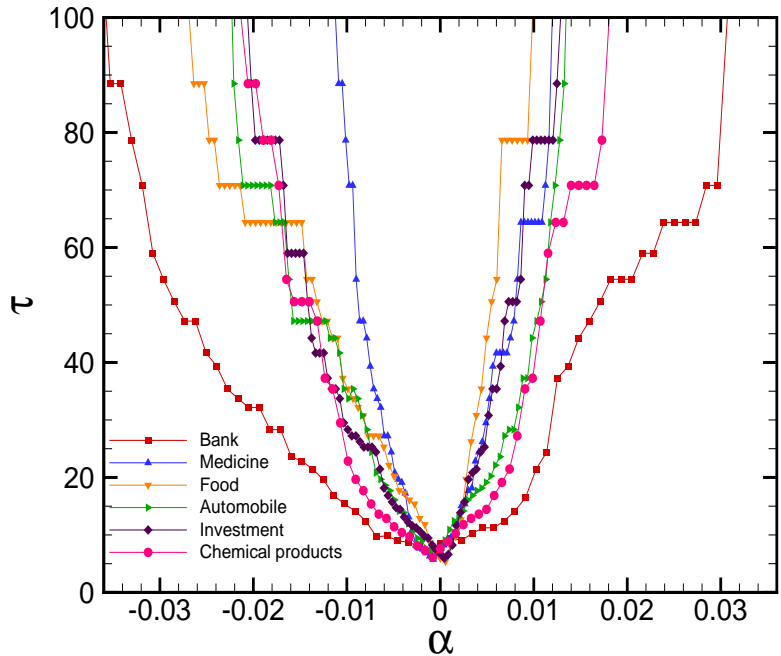}
\caption{Comparison of the waiting time of some TEPIX subgroup
indices.}\label{fig4}
\end{figure}
\begin{figure}[t]
\includegraphics[width=18cm,height=8cm,angle=0]{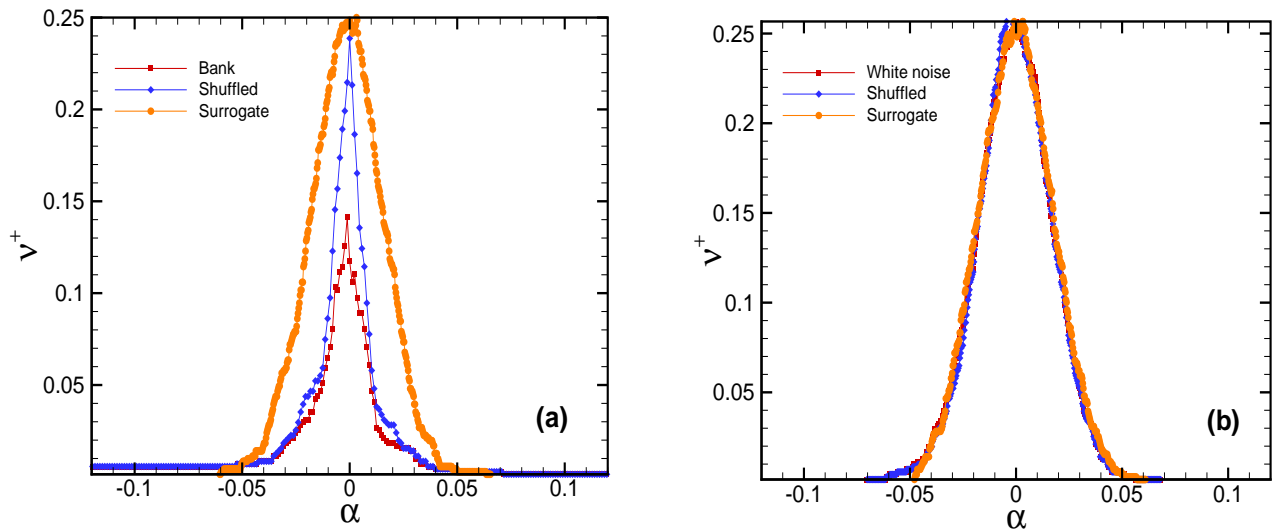}
\caption{Typical comparison of the positive Level Crossings of bank
index, its shuffled \& surrogate with a white noise (the standard
deviations of the curves are the same).}\label{fig5}
\end{figure}
\begin{figure}[t]
\includegraphics[width=18cm,height=8cm,angle=0]{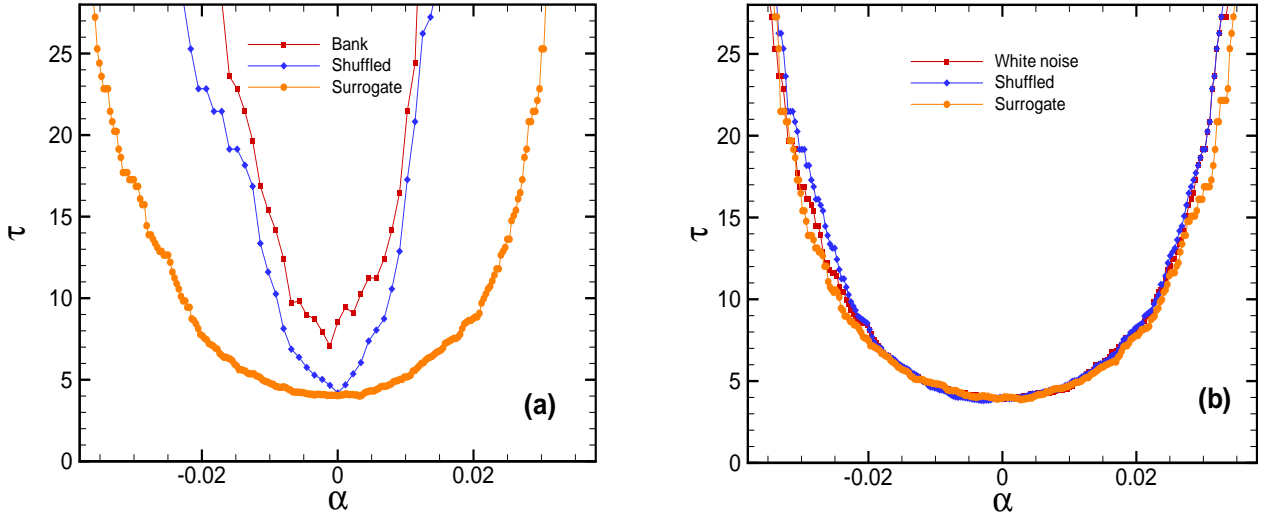}
\caption{Typical comparison of the waiting times of bank index, its
shuffled \& surrogate with a white noise (the standard deviations of
the curves are the same).}\label{fig6}
\end{figure}
\begin{figure}[t]
\includegraphics[width=12cm,height=10cm,angle=0]{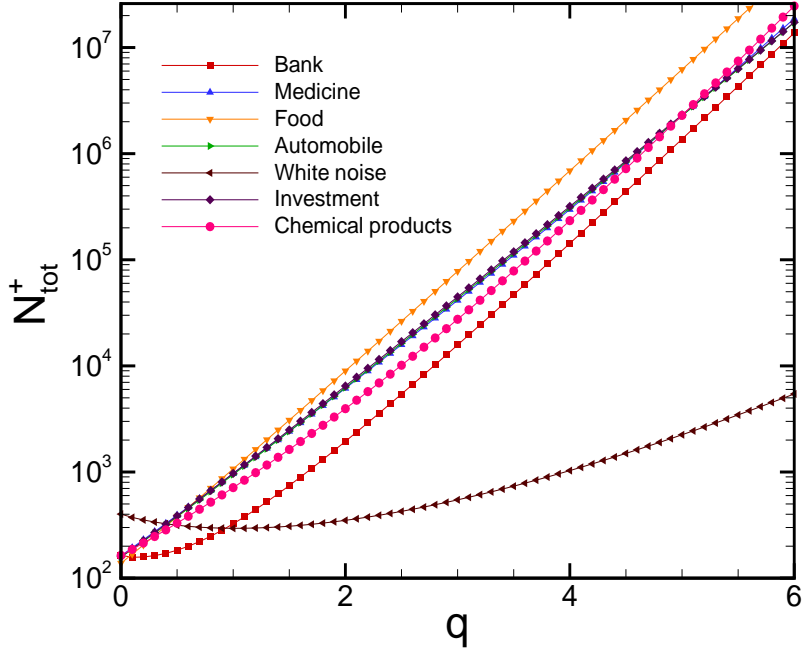}
\caption{Generalized total number of crossings with positive slope
$N^{+}_{tot}(q)$ for some TEPIX subgroup indices.}\label{fig7}
\end{figure}
\begin{figure}[t]
\includegraphics[width=18cm,height=16.5cm,angle=0]{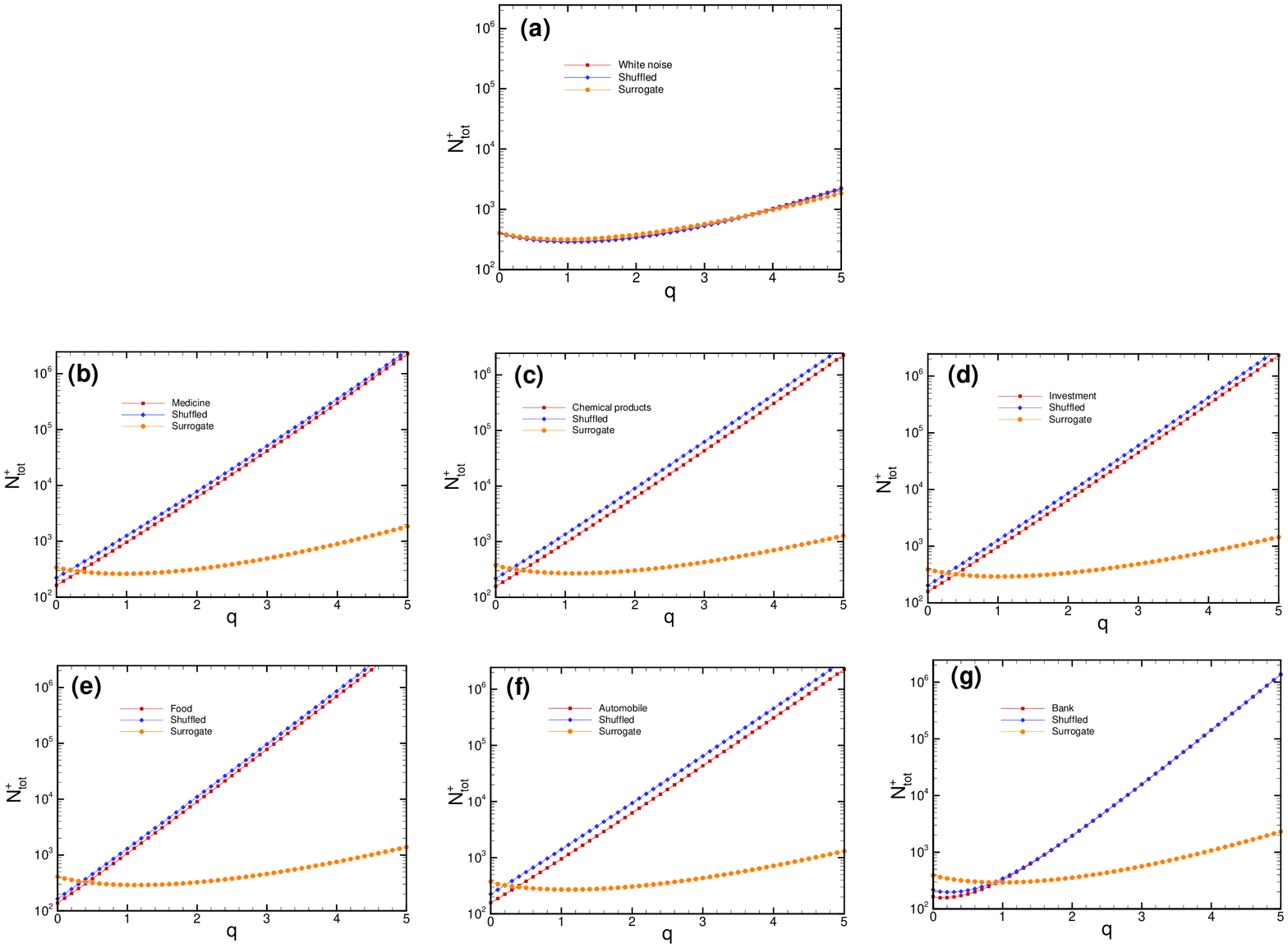}
\caption{Comparison of generalized total number of Level Crossings
with positive slope $N^{+}_{tot}(q)$ of (a) White noise, (b)
Medicine, (c) Chemical products, (d) Investment, (e) Food, (f)
Automobile and (g) Bank with their shuffled and
surrogate.}\label{fig8}
\end{figure}

Actually what we really mean is that there will be high probability
of a crossing in interval $\Delta t$ if these conditions are
satisfied \cite{Tabar1}.

In order to determine whether the above conditions are satisfied at
any arbitrary location $t$, we must find how the values of $y= r -
\bar r $ and $ y ^{\prime}= \frac{ \Delta y }{\Delta t}$ are
distributed by considering their joint probability density
$p(y,{y}^{\prime})$. Suppose that the level $y=\alpha$ and interval
$\Delta t$ are specified. Then we are only interested in values of
$y < \alpha$ and values of ${y}^{\prime}=(\frac{\Delta y}{\Delta t})
> \frac{\alpha-y}{\Delta t}$ , which means that the region between
the lines $y=\alpha$ and ${y}^{\prime}= \frac{\alpha-y}{\Delta t}$
in the plane ($y,{y}^{\prime}$). Hence the probability of positive
slope crossing of $y=\alpha$ in $\Delta t$ is \cite{Tabar1}:
\begin{equation}
\int_{0}^{\infty} \Delta{y}^{\prime}\int_{\alpha-{y}^{\prime}\Delta
t}^{\alpha} \Delta y p(y,{ y}^{\prime}).
\end{equation}

When $\Delta t\rightarrow 0$, it is legitimate to put

\begin{equation}
p(y,{y}^{\prime})=p(y=\alpha,{y}^{\prime}).
\end{equation}
Since at large values of $y$ and ${y}^{\prime}$ the probability
density function approaches zero fast enough, therefore eq.(6) may
be written as \cite{Tabar1}:

\begin{equation}
\int_{0}^{\infty} d{y}^{\prime}\int_{\alpha-{y}^{\prime}d
t}^{\alpha} d y p(y=\alpha,{y}^{\prime})
\end{equation}
in which the integrand is no longer a function of $y$ so that the
first integral is just: $\int_{\alpha-{y}^{\prime}d t}^{\alpha} d y
p(y=\alpha,{y} ^{\prime})=p(y=\alpha,{y}^{\prime}){y}^{\prime}d t $,
so the probability of slope crossing of $y=\alpha$ in $d t$ is equal
to \cite{Tabar1}:

\begin{equation}
d t\int_{0}^{\infty} p(\alpha,{y}^{\prime}){y}^{\prime}d{y}^{\prime}
\end{equation}
in which the term $p(\alpha,{y}^{\prime})$ is the joint probability
density $ p(y,{y}^{\prime})$ evaluated at $y=\alpha$.

We have said that the average number of positive slope crossings in
scale $T$ is $\nu_{\alpha}^{+} T$, according to $(4)$. The average
number of crossings in interval $d t$ is therefore
$\nu^{+}_{\alpha}d t$. So, average number of positive crossings of
$y=\alpha$ in interval $d t$ is equal to the probability of positive
crossing of $y=a$ in $d t$, which is only true because $d t$ is
small and the process $y(t)$ is smooth so that there cannot be more
than one crossing of $y=\alpha$ in time interval $d t$, therefore we
have $\nu_{\alpha}^{+}d t=d
t\int_{0}^{\infty}p(\alpha,{y}^{\prime}){y} ^{\prime}d{y}^{\prime}$,
from which we get the following result for the frequency parameter
$\nu_{\alpha}^{+}$ in terms of the joint probability density
function $p(y , {y}^{\prime})$
\begin{equation}\label{level}
\nu_{\alpha}^{+}=\int_{0}^{\infty}p(\alpha,{y}^{\prime}){y}^{\prime}d{y}
^{\prime}.
\end{equation}

Some authors have used other forms for $\nu_\alpha^+$ which are as
follows \cite{oil}:
\begin{eqnarray}
\nu_\alpha^+= P(y_i>\alpha,y_{i-1}<\alpha)
\end{eqnarray}
\begin{eqnarray}
\nu_\alpha^+ &=& \int_{-\infty}^\alpha\int_\alpha^\infty
P(y_i,y_{i-1})d y_i d y_{i-1} \cr \nonumber\\
&=& \int_{-\infty}^\alpha\int_\alpha^\infty P(y_i|y_{i-1})P(y_{i-1})
d y_i d y_{i-1}\;,
\end{eqnarray}

Let us also use the quantity $N_{tot}^{+}(q)$ as \cite{jafari1}
 \begin{equation}\label{ntq}
N_{tot}^{+}(q)=\int_{-\infty}^{+\infty}\nu_{\alpha}^{+} |\alpha -
\bar{\alpha}|^{q} d\alpha.
\end{equation}
where zero moment (with respect to $\nu_{\alpha}^{+}$) $q=0$, shows
the total number of crossings for price returns with positive slope.
The moments $q<1$ give information about the frequent events while
moments $q>>1$ are sensitive for the tail of events.

LC analysis is very sensitive to correlation when the time series is
shuffled and to probability density functions (PDF) with fat tails
when the time series is surrogated. To study the effects of
correlations and probability density functions (PDF), we have
evaluated $N_{sh}^{+}$ (which is the total number of positive-slope
crossings of the height fluctuation series when it is shuffled) and
$N_{su}^{+}$ (which is the total number of positive-slope crossings
of the series when it is surrogated). The shuffling and surrogating
procedures are explained in the following:

\subsection{Shuffling procedure}

A celebrated theorem of Aldous, Bayer, and Diaconis asserts that it
takes $\frac{3}{2}\log_{2} n $ riffle shuffles to randomize a deck
of n cards, asymptotically for large n, and that the randomization
occurs abruptly according to a "cutoff phenomenon." Shuffling by
random transpositions is one of the simplest random walks on the
symmetric group: given n cards in a row, at each step two cards are
picked uniformly at random and exchanged. This shuffle was precisely
analyzed in 1981 \cite{Ram}. We have used random transpositions for
shuffling our data. The long range correlations are destroyed by the
shuffling procedure. As it will be pointed, in liquid markets,
correlations of returns are small and in inefficient markets
correlations are large. Hence, by comparing the original returns
with the shuffled ones we can obtain the magnitude of correlations
in the market and this can help us gain useful information about the
market.

\subsection{Surrogating procedure}

Another procedure that is used for obtaining valuable information
about the time series is surrogating procedure. In the surrogate
method "surrogates" are generated by replacing the true phases with
a set of pseudo independent distributed uniform $(-\pi,+\pi)$
generated by any good pseudorandom uniform subroutines
\cite{surrogate}. The phase of the discrete Fourier transform
coefficients of time series are replaced with a set of
pseudo-independent distributed uniform $(-\pi,+\pi)$ quantities
generated by any good pseudorandom uniform subroutines. The
correlations in the surrogate series do not change, but the
probability function changes to Gaussian distribution
\cite{surrogate,the92,the93,the97,sun,Bach,mahsa}. The main
objective is to provide a kind of baseline or control against which
the original data can be compared. The physical idea behind the
surrogates method is that a nonlinear operation on a stationary
random forcing process generates cross frequency coupling between
complex amplitudes. The discrete Fourier transform (DFT) of the
observed time series data is computed and then the phases of each
complex amplitude of the DFT are replaced with independently
distributed artificial uniform $(-\pi,+\pi)$ variates. The altered
DFT is then inverse Fourier transformed to generate a surrogate time
series. The randomization ensures that any phase coupling, and thus
signs of nonlinearity, is destroyed in the surrogates.

\section{The strategy to develop the Market}

In general, a successful privatization program requires
institutional changes that contribute significantly to the
strengthening of the legal framework underlying equity investment.
However, private control and policy reforms must be maintained
during any political backlash. As a consequence, actual progress of
privatization builds up confidence over time and this will lead to
market deepening, investment and trading. This may explain why
privatization may be contemporaneous or even precede successful
stock market development. Alternative benefits of successful
privatization are improved risk sharing and increased liquidity and
activity of the market \cite{perotti,jafari1,mahsa}. One of the main
methods to develop privatization, is entering a new stock to the
markets for arising competition. But attention to the capability of
the markets to accept a new stock is substantial. Without
considering the above statement, it is possible to reduce the
market's efficiency. In other words, introduction of a new stock to
the market usually decreases the stage of development and activity
and increases the risk. It has been shown that inefficient markets'
stage of development and activity are lower and their risk is higher
than efficient markets \cite{jafari1,mahsa,Pagano}.

The reason why, in very efficient markets of equities and currency
exchanges correlations of returns are extremely small, is because
any significant correlation would lead to an arbitrage opportunity
that is rapidly exploited and thus washed out. Indeed, the fact that
there are almost no correlations between price variations in
efficient markets can be understood from simple calculation
\cite{Sornette,ro}. In other words, liquidity and efficiency of
markets control the degree of correlation, that is compatible with a
near absence of arbitrage opportunity. It is important to consider
that, the more intelligent and hard working the investors, the more
random is the sequence of price changes generated by such a market.

In the following subsections we try to explain more about activity,
stage of development, risk and waiting times which can characterize
the market and should be considered before privatization.

\subsection{Activity}

One of the parameters that should be considered before privatization
is activity. When there is no sell and buy in the market, the prices
are fixed without any fluctuation. One should pay attention that
fluctuation is calculated with respect to the trading volume. With
increasing the trading volume, the fluctuation decreases. Indeed,
fluctuation is the sign of existence of sell and buy that is called
activity \cite{mahsa}. High fluctuation could also increase the
risk. The process of buy and sell or activity is a positive
parameter which could be effective in determining the real price of
the stock and in correctly distributing the wealth. It is obvious
that by increasing activity, liquidity increases which can stimulate
investors to enter their short-term investment to the market that
can again increase (improve) activity. The liquidity of a product
can be measured as how often it is bought and sold. Liquidity is
characterized by a high level of trading activity.

\subsection{Stage of development}

Based on recent research for characterizing the stage of development
of markets \cite{Payam,mahsa,Matteo1,Matteo2}, it is shown that the
Hurst exponent (H) has sensitivity to the degree of development of
the market. In liquid markets, correlations of returns are small
because existence of any information in the market would lead the
investors to get use of it and thus washed out. In contrast, if
there is no correlation between price variations or if markets are
perfectly efficient, the return on gathering information is nil.
Therefore, there would be little reason to trade and markets would
eventually collapse \cite{gross}.

One of the important points in market development, is acceptability
of the market development by investors. In emerging markets, there
exist correlations and information (high value of Hurst exponent)
which can stimulate the investors to gain benefits. If the
development is accepted by investors, it will be a successful
program. In this case, the market parameters improve (higher
activity and stage of development and lower risk). In most
developing countries and even in some developed countries, "they
privatize just for privatization" and not for promoting efficiency,
improving the operations of markets, reduction in public debt, etc
and in this case there will occur an unsuccessful privatization.
This happens because they have not enough attention to the above
point and so these markets have not the capability of accepting the
new stocks and the development of privatization. This suggests that
by considering the current situation, the rate of development and
privatization should be controlled. Here, we could ask a question:
why the investors do not get use of the present information besides
the development?

 \textbf{A:} Lack of "liquid investment": This means that the worth of
the introduced stocks is more than the worth of the liquid
investment. In this case liquidity of the stocks decreases. In other
words, exchange of the stocks is reduced and so the liquidity of the
market falls and there will be an inefficient and frozen market. In
this case privatization is not a successful program for development
of the market. In other words, before privatization, attention to
the worth of the introduced stocks in comparison with the liquid
investment is necessary.

\textbf{B:} Lack of financial security and high risk: This factor
ruins the motives of investors for investment. This is one of the
main problems in developing countries.

\textbf{C:} "Frozen investment": In some countries, there is not
enough study or proper management program for privatization and
instead of using privatization for the use of potential intelligence
and investment, they privatize for privatization. For example,
consider the case that government distribute some stocks between
people who did not want to be investors on their own. Hence they do
not participate in the exchange of stocks and currency and they form
the "frozen investment". Thus they cannot increase the intelligence.
These are the stocks that were frozen under the name of the
government and now only the name is changed. While the aim of
privatization is not the apparent change of the names of the
investors.

Thus, for a successful privatization program these three parameters
should also be considered. We intend to study development of the
market from this aspect by using Level Crossing analysis.

LC with the power of correlation detection, is a useful tool to find
the stage of development of markets \cite{jafari1,mahsa}. It is
known that inefficient markets have long-range correlation. This
sensitivity of LC to the market conditions provides a new and simple
way of empirically characterizing the development of financial
markets. This means that, in mature markets the total positive slope
crossings, $N^{+}_{tot}$ is fixed or decreases under shuffling
effectively, while in emerging markets, it is increased. Recently,
many works have focused on the stage of development of the market
\cite{Matteo1,Matteo2,Payam,jafari1,mahsa,Garch}. As far as stock
markets are concerned, the Hurst exponents show remarkable
differences between developed and emerging markets. Di Matteo et al
(2003) and (2005) found that the emerging markets have $H>0.5$
whereas the developed ones have $H\leq 0.5$
\cite{Matteo1,Matteo2,Payam}.

\subsection{Risk}

Nowadays, the importance of risk is clear for everyone. As many
researches have shown, many financial crisis such as Asian crisis
during 1997-1998 have been related to the lack of suitable risk
management. Indeed, risk is one of the important subjects that has
been considered by most of the financial organizations e.g. banks,
assurance companies etc. In other words, precise recognition of the
financial markets and the means to maintain the stability in these
markets, is one of the crucial ways to preserve the economic growth
of the country. Risk management is one of the main tools for
stability of financial markets. In fact, market risk is resulted
from the high fluctuation (low frequency regime) in the prices of
assets. Assets could be in the form of cash, stocks, lands, gold
etc. All of these could have fluctuation in their prices and this is
the main result to produce risk. Thus, assessing and pricing the
risk properly is of substantial importance. Financial and credit
risks \cite{risk1,risk2,risk3,risk4} play the main roles in
bankruptcy of the financial markets. These continuous crisis due to
the financial risk have better shown the necessity of attention to
the management of risk.

In fact, most of the common methods for risk measurement is based on
Value at Risk (VaR) \cite{risk1,risk2,risk3}. VaR is a downside risk
measure, meaning that it typically focuses on losses. Most of the
common methods (ARCH models, variance-covariance, riskmetrics ...)
are based on variance and the kinds of it. We want to look at risk
from another point of view because of its importance. In fact, small
fluctuation in the return of the prices have contribution in the
variance and so in risk. But we mention that small fluctuation
demonstrate the activity of the market which is a positive
parameter. In a market with no buy and sell there is no fluctuation
in the price. In this case the variance is zero and there is no
risk. It means that we should consider more contribution from larger
fluctuation. But we should pay attention to the point that risk is
not a totally negative concept. Although the investors could hold
their money in the banks without risk, they prefer to invest it in
the markets that could have risk. This operation depends on the risk
that an investor will take and this is due to two main factors: 1)
The worth of his investment and 2) Human psychology. Indeed, in
average, the more his investment, the more risk he will take. One of
the market factors is knowledge of distribution of the investors and
their investment. It is clear that human psychology is important and
this is related to social sciences.

\subsection{Investment horizons}

To speak about other aspects of the development of markets, we can
mention investment horizons. By investment horizons we mean the
expectation time for a specified benefit. The quantitative method
that is used is inverse statistics which was introduced by Johansen
et al \cite{Johansen1,Johansen2,Johansen3}. This method has been
formed by the main idea that the current prices are the results of
the future expectation. In other words, in response to the question
of "how to price in the available information?" for the case of a
stock, one must consider how the available information affects
future earnings of the company. This introduces some ambiguity as
not only do peoples expectations to a largely unpredictable future
differ, but so do their strategies. But how we can quantify the
peoples expectations? One of the important parameters that affect
peoples expectations is the expectation time which is defined as the
specified time interval that one should wait for a specific change
in the price value. The answer to this question can adjust the
investors expectations with respect to for example investment
horizons. This expectation time can help the investors to invest
their money by considering the risk they can take and the fact that
how long they intend to hold their investment before taking any
profit. The risk that people are willing to take is a subject of the
human psychology and social sciences. This can suggest that
economics and financial problems are coupled with human sciences.

\section{application}

To study the market development, we have analyzed some selected
TEPIX subgroup indices using Level Crossing analysis (Fig. 2). These
indices are Bank, Medicine, Food, Automobile, Investment, Chemical
products. The data are taken from Tehran Securities Exchange
Technology Management Company (TSETMC) \cite{tsetmc}. Log return
time series, $r(t)$ of them from the same time interval: 3 January
2005 to 1 March 2008 have been analyzed. Data have been recorded at
each trading day. In addition, we have compared the results with a
white noise as a standard reference.

According to Eq. \ref{level} Level Crossing, $\nu^{+}_{\alpha}$, is
calculated for the indices. Figure \ref{fig3} shows a comparison of
$\nu^{+}_{\alpha}$ for TEPIX subgroups as a function of level
$\alpha$. It is clear that $\nu^{+}_{\alpha}$ scales inversely with
time, so $\tau(\alpha)=\frac{1}{\nu^{+}_{\alpha}}$ is a waiting time
which the level $\alpha$ will be observed again. Figure \ref{fig4}
shows a comparison of $\tau(\alpha)$ for TEPIX subgroups. These
figures show that the behavior of various indices is different.

Table \ref{Tb1} shows the waiting time for meeting various levels
from high frequency to low frequency (tails) regimes:
($\tau(\alpha=0)$) which is the mean of the levels,
($\tau(\alpha=\pm0.005)$, $\tau(\alpha=\pm0.01)$ and
$\tau(\alpha=\pm0.02)$ for all selected indices of TEPIX subgroups.
The time interval $\tau(\alpha=0)$ of Bank index is the largest
($8.53$) and the time interval $\tau(\alpha=0)$ of Food index is the
smallest ($5.95$) days which are in the same order. This means that
in average, it takes more time for meeting the mean of the levels
with positive slope for Bank index than Food index. In other words,
the Food index lives more in the level $\alpha=0$. As it is seen in
the table \ref{Tb1}, there exists an asymmetric behavior when we are
moving towards the tails (high levels) and this behavior is in such
a way that there is a tendency towards the mean, except the index of
Medicine. In all indices except Medicine index, the waiting time
interval of right is larger than left tails. This means that when we
are in the levels larger than mean we should wait more time for
meeting this level with positive slope, than for the same opposite
level. According to table \ref{Tb1}, in average, the waiting time
interval in the tails for Bank index is the smallest, which means
that it is financially motivated to absorb capital in this index.

The area under the curves of $\nu^{+}_{\alpha}$ shows the total
positive Level Crossings, $N^{+}_{tot}$ which represents the
activity of the market. As it is seen in table \ref{Tb2}, Food index
has the smallest activity ($138.17$) and Bank index has the largest
one ($164.28$). This should be noted that the values of the
$N^{+}_{tot}$ are reported for this time scale (3 Jan. 2005 to 1
Mar. 2008). As it is reported in table \ref{Tb1}, the waiting time
of the Bank index is the smallest in various levels which means it
is more active. Activity or the process of buy and sell is a
positive parameter which could be effective in determining the real
price of the stock and in correctly distributing the wealth. By
increasing activity, liquidity increases.

Figures \ref{fig5} and \ref{fig6} show typical comparison of the
positive Level Crossings and waiting times of an index (Bank index)
and white noise with their shuffled and surrogate, respectively. In
order to better compare the results, standard deviation of the white
noise is chosen to be the same as the Bank index. It should be noted
that, in these figures the standard deviation of the shuffled and
surrogate are kept the same as the original ones. Considering that,
for a white noise, the Level Crossing curves do not change after
shuffling and surrogating processes, any deviation from this
behavior shows existence of statistical information and departure
from the white noise. As it is seen in figure \ref{fig5}(a), when we
are shuffling and surrogating the returns of Bank index, the Level
Crossing curve changes. Hence, the area under the curves of the
original and its shuffled and surrogate differs. As it is reported
in table \ref{Tb2}, $N^{+}_{tot}$, $N_{sh}^{+}$ and $N_{su}^{+}$ of
the Bank index are $164.28$, $218.31$ and $389.83$, respectively. As
it is seen in figure \ref{fig6}, the waiting time for the Bank index
and its shuffled and surrogate are shown. The waiting time for the
surrogate of the Bank index grows much slower than the original and
the shuffled series. This represents that the Bank index is far from
the Gaussian function. This shows the existence of risk that will be
studied later in this section. As it is seen in the figures
\ref{fig5} and \ref{fig6}, the difference between the surrogate of
Bank index and white noise is much smaller than the difference
between the shuffled Bank index and the white noise which lead us to
the conclusion that the contribution from the PDF is larger than the
contribution from the correlation. This is because the index is
fat-tailed. This behavior is in good agreement with the results of
the table \ref{Tb2}. As it is reported in this table, comparison
between the relative differences for the Bank index (
$|N_{sh}^{+}-N_{tot}^{+}|/N_{tot}^{+}$ and
$|N_{su}^{+}-N_{tot}^{+}|/N_{tot}^{+}$ which are $0.33$ and $1.37$
respectively) will reveal that the contribution from the PDF is
larger. The results of other indices are listed for better
comparison.

Since activity, $N^{+}_{tot}(q=0)$ is very sensitive to correlation,
it changes when the time series is shuffled so that the correlation
disappears. When the relative difference of the shuffled and the
original series, $|N_{sh}^{+}-N_{tot}^{+}|/N_{tot}^{+}$ is positive,
the series is correlated and when this relative difference is
negative, the series is anti-correlated. This behavior may not be
seen when calculated directly by standard models. One of the
advantages of Level Crossing method is that no scaling feature is
explicitly required. Thus, by comparing the difference between
$N^{+}_{tot}(q=0)$ and $N^{+}_{sh}(q=0)$ (after shuffling), the
stage of development of markets can be determined. Smaller relative
difference denotes larger stage of development. The results of this
relative difference is listed in table \ref{Tb2}, which the minimum
is $0.19$ for Food index and the maximum is $0.38$ for the
Automobile index. These results show that the Medicine index is the
index with highest stage of development and the stage of development
of Automobile index is the lowest.

Also, activity, $N^{+}_{tot}(q=0)$, is sensitive to deviation of PDF
from normal distribution. Thus, it changes when the time series is
surrogated so that the PDF changes to a Gaussian one. When the
relative difference of the surrogate and the original series,
$|N_{su}^{+}-N_{tot}^{+}|/N_{tot}^{+}$ is positive, the series is
fat-tailed and when this relative difference is negative, Gaussian
distribution is wider than the series. Thus, by comparing the
difference between $N^{+}_{tot}(q=0)$ and $N^{+}_{su}(q=0)$ (after
surrogating), the contribution of sudden changes in activity can be
determined. Larger relative difference denotes larger risk. The
results of this relative difference is listed in table \ref{Tb2},
which the minimum is $339.56$ for Medicine index and the maximum is
$408.60$ for the Food index. All the relative changes are positive
which means that all the indices are fat-tailed. Fat tails of a
distribution refers to a much larger probability for large price
changes than what is to be expected from the random walk or Gaussian
hypothesis and they imply additional risk.

For a better comparison, Hurst exponent which was obtained by using
the detrended fluctuation analysis (DFA) method \cite{Peng1,Peng2},
are reported in table \ref{Tb2} too. As it is seen they are in
agreement with the results of the Level Crossing method. But it
should be noted that errors in evaluating the exponent is more.
Inefficient markets are associated with high value of Hurst exponent
and developed markets are associated with low value of the exponent.
In particular, it is found that all emerging markets have Hurst
exponents larger than $0.5$ (strongly correlated) whereas all the
developed markets have Hurst exponents near to or less than 0.5
(white noise or anti-correlated). As reported
\cite{Matteo1,Payam,jafari1}, at one end of the spectrum, there are
stocks like Nasdaq 100 (US), $S\&P500$ (US), Nikkei $225$ (Japan)
and so on. Whereas, at the other end, there are TEPIX, the
Indonesian JSXC, the Peruvian LSEG, etc. We notice that TEPIX belong
to the emerging markets category and it is far from an efficient and
developed market.

When we apply Eq. \ref{ntq} for small $q$ regime, high frequency
events ($\alpha = 0$) are more significant, whereas in the large $q$
regime, low frequency (the tails) is more significant. Figure
\ref{fig7} compares the generalized total number of positive slopes
Level Crossings, $N^{+}_{tot}(q)$ of different indices and a white
noise. In this figure, for better comparison, we choose the same
variance for all of them. As it is seen in the figure,
$N^{+}_{tot}(q)$ for the Bank index is below the other indices for
$q>>1$ and the Food index is above them. This means that in these
indices, Food index is higher in risk and Bank is lower. This figure
shows how $N^{+}_{tot}(q)$ for the indices are deviated from a white
noise. To know the reason of this deviation we compare the behavior
of $N^{+}_{tot}(q)$ for each index with its shuffled and surrogate
in figure \ref{fig8}. White noise behavior has been plotted for a
better decision. It is clear that $N^{+}_{tot}(q)$ of white noise
and its shuffled and surrogate have the same behavior because a
white noise is an uncorrelated series with normal distribution. The
surrogated $N^{+}_{tot}(q)$ for all the indices shows little
difference from the white noise. This means that PDF leads this
deviation in our indices. $N^{+}_{tot}(q)$ for the moments $q>>1$ is
sensitive to the tail of events which is the sudden changes. As it
is shown in the figure for $q>>1$, the largest difference between
$N^{+}_{tot}(q)$ for the original and the surrogate series is for
Food index and the smallest difference is for the Bank index. This
means that Food index is more risky than other indices and the risk
of the Bank index is the smallest. Another point that should be
noted, is that while for $q<1$, $N^{+}_{tot}(q)$ is different for
Bank index and its shuffled, this difference could not be seen for
$q>>1$. This shows that the existence of correlation in high
frequency regime is more than the low frequency regime.

Considering all of the above discussions and results, we notice that
TEPIX and most of its subgroups belong to the emerging markets
category and they are far from efficient and developed markets.
Tables \ref{Tb1} and \ref{Tb2} compare the properties of these
subgroup indices. Knowledge of distribution of the investors and
their investment could be important. In general, the stage of
development is substantial for developing the market. Higher stage
of development means that this market is closer to an efficient
market. Thus, the capability to accept a new stock is more. When
there exist investors with limited investment (less risk-taking
investors) higher activity and lower risk could be more important.
The Food index with highest stage of development is closer to an
efficient market but with its lower activity and higher risk for its
development, we need more risk-taking investors (the more investment
one has, the more risk he will take). In contrast, Bank with highest
activity and lowest risk has the middle place in the stage of
development. Thus, if investors with limited investment exist, these
two parameters are more important. Hence, for development and
privatization, knowing the distribution of the investors and their
investment and also studying the human psychology is necessary.

\begin{table}
\caption{\label{Tb1} The values of waiting time, $\tau(\alpha)$ for
different levels, $\alpha$ for some TEPIX subgroup indices.}
\medskip
\begin{tabular}{|c|c|c|c|c|c|c|c|}
  \hline
  Index & $\tau(\alpha=0)$ & $\tau(\alpha=-0.005)$ &  $\tau(\alpha=0.005)$&$\tau(\alpha=-0.01)$ & $\tau(\alpha=0.01)$ & $\tau(\alpha=-0.02)$ & $\tau(\alpha=0.02)$ \\\hline
  Bank  & 8.53 & 9.38& 11.24&15.39 &21.45 & 32.18 &54.46  \\\hline
  Medicine & 7.87 &22.47  & 31.47 &78.68 & 64.35 &353.98  & 236.02  \\\hline
  Chemical products &7.61&11.42 &14.45&29.50 &59.00&88.50 &101.14 \\\hline
  Automobile &7.30 &16.09 &19.14 &35.40  &47.19  &70.80&176.99 \\\hline
  Investment  &6.74& 15.73& 30.78& 33.70& 78.68& 56.64 &236.02\\\hline
  Food  & 5.95 &20.23 & 44.27 & 37.26 &  101.11& 64.36 &236.02  \\\hline
\end{tabular}
\end{table}

\begin{table}
\caption{\label{Tb2} The values of total number of crossings with
positive slope for some TEPIX subgroup indices, $N^{+}_{tot}(q=0)$
(activity), their shuffled, surrogate, their relative differences
between original data \& their shuffled (stage of development) and
surrogate (deviation from a normal distribution) and comparison of
them with Hurst exponents.}
\medskip
\begin{tabular}{|c|c|c|c|c|c|c|}
  \hline
  Index & $N_{tot}^{+}$ & $N_{sh}^{+}$ & $N_{su}^{+}$ & $|N_{sh}^{+}-N_{tot}^{+}|/N_{tot}^{+}$ & $|N_{su}^{+}-N_{tot}^{+}|/N_{tot}^{+}$& $H$ \\\hline
  Automobile & 157.60& 217.85 & 374.19 &0.38 &1.37 &0.71 $\pm$ 0.02 \\\hline
  Medicine & 162.57& 222.49& 339.56 & 0.37 & 1.09 & 0.64 $\pm$ 0.02  \\\hline
  Bank & 164.28 & 218.31 & 389.83 & 0.33 & 1.37 & 0.61 $\pm$ 0.02  \\\hline
  Chemical products &163.89 & 212.61& 385.44&0.30 &1.35 &0.60 $\pm$ 0.02 \\\hline
  Investment  &158.87& 203.67 & 390.56 & 0.28 &1.46 &0.56$\pm$ 0.02 \\\hline
  Food & 138.17& 164.20 & 408.60 & 0.19& 1.96 &0.55  $\pm$ 0.02 \\\hline

  \end{tabular}
\end{table}

\section{Conclusion}

In this chapter, we have studied some concepts, which are important
to develop and privatize markets by Level Crossing approach. We have
calculated some parameters of indices which are effective parameters
in development and privatization of markets. These parameters
include: activity, stage of development, risk and investment
horizons (which depends on waiting time). The subgroups with higher
activity and stage of development and lower risk and appropriate
investment horizons (the waiting time interval that one should wait
for a specific change in the price value) are more suitable for
development and privatization than other subgroups. This method is
based on stochastic processes which should grasp the scale
dependency of any time series in a most general way. We have applied
this method to $6$ selected subgroup indices (Bank, Medicine, Food,
Automobile, Investment, Chemical products) of Tehran stock market
and compared their properties.


\end{document}